\documentclass[12pt,preprint]{aastex}

\newcommand{\eb}{\begin{equation}}
\newcommand{\ee}{\end{equation}}

\shorttitle{Passage and entrapment of terrestrial planets in spin-orbit resonances}
\shortauthors{Makarov}
\begin{document}

\title{How terrestrial planets traverse spin-orbit resonances: A camel goes through a needle's eye} 
\author{Valeri V. Makarov}
\affil{United States Naval Observatory, 3450 Massachusetts Ave. NW, Washington DC, 20392-5420}
\email{vvm@usno.navy.mil}

\begin{abstract}
The dynamical evolution of terrestrial planets resembling Mercury in the vicinity of
spin-orbit resonances is investigated using comprehensive harmonic expansions of the tidal
torque taking into account the frequency-dependent quality factors and Love
numbers. The torque equations
are integrated numerically with a small step in time, includng the oscillating triaxial 
torque components but neglecting the layered structure of the planet and assuming a zero
obliquity.
We find that a Mercury-like planet with its current value of orbital eccentricity (0.2056)
is always captured in the 3:2 resonance. The probability of capture in the higher 2:1 resonance
is approximately 0.23. These results are confirmed by a semi-analytical estimation of
capture probabilities as functions of eccentricity for both prograde and retrograde
evolution of spin rate. As follows from analysis of equilibrium
torques, entrapment in the 3:2 resonance is inevitable at eccentricities between $0.2$
and $0.41$.
Considering the phase space parameters at the times of periastron, the range
of spin rates and phase angles, for which an immediate resonance passage is triggered, is very
narrow, and yet, a planet like Mercury rarely fails to align itself into this state of unstable equilibrium
before it traverses the 2:1 resonance.
\end{abstract}

\keywords{planets and satellites: dynamical evolution and stability --- celestial mechanics --- 
planet-star interactions}
\section{Introduction}
\label{firstpage}
\label{Introduction}
Planets of terrestrial type with solid mantles are subject to triaxial and tidal torques exerted
by the host star. A planet just formed from the protoplanetary disk rotates at a much higher rate than the
rate of its orbital motion. In the course of millions to billions of years, the secular terms of the tidal torque cause the planet to spin down. The tidal bulges moving across the planet at different
frequencies result in a gradual loss of kinetic energy through friction and heating.
The energy dissipation rate is normally so slow, that most of the major planets in the Solar system
still rotate faster than they revolve around the Sun, with the exception of Venus with its slow
retrograde rotation and Mercury,
which is in the 3:2 spin-orbit resonance \citep{pett}. Presumably, the planet traversed a number of
higher-order resonances before it reached this state. The ultimate, and the most stable, state for a rotating
planet subject to tidal forces is the 1:1 resonance, when the rotation rate is equal to the orbital rate, and the planet is always pointing with its most elongated dimension toward the star. The dissipation of
the energy of rotation also diminishes the obliquity of the planets equator, gradually aligning
the axes of rotation and of the orbit. Mercury's dynamical evolution has been much faster than that
of other solar planets, because 1) it is closer to the Sun; 2) its orbital eccentricity varied in a relatively
wide range and has been higher than the eccentricity of the other close-in planets. 
The importance of eccentricity for Mercury's chaotic evolution
was emphasized by \citet{corla04,corla09}, who revised upward the original estimate of the probability
of capture to the present 3:2 resonance at 7\% \citep{gold}.

Mercury can serve as a good model for smaller mass, rocky exoplanets, especially those orbiting
M dwarfs in their habitable zones. Although the exact rheology of exoplanets will remain a matter
of speculation for the foreseeable future, it seems reasonable to adopt the parameters and models
obtained for the Earth and the Moon. The objective of this paper is to investigate the circumstances
of the transition of a Mercury-like planet with an Earth-like rheology through a high order spin orbit resonance. In this case, the widely accepted approximations for the value of tidal torque in the
vicinity of a resonance are not applicable. Furthermore, the oscillatory terms of the
force can not be neglected. We employ high-order expansions of the torque
in harmonics of tidal frequency and powers of eccentricity, and a relation for the Love number
as a function of tidal frequency in terms of real and imaginary compliances(\S 2). 
The resulting differential equation of second order, which
includes both the tidal and triaxial torque components, is integrated with a step much smaller than
the period of rotation, with the current best estimates for Mercury (\S 3). With the current
value of eccentricity ($e=0.20563$), the planet traverses the 2:1 resonance with an estimated
probability of 0.77, and is
always captured in the 3:2 resonance. The transition is very fast, accompanied by a significant step-down
in the average rotation rate. In \S \ref{con.sec}, more integrations are performed with 
the initial rate set to
exactly the 2:1 resonance and with various initial phase angles. It is revealed that such a planet
inserted in resonance almost always stays in it. The actual passage through the resonance can only occur
through a tiny area of the phase space around a phase angle of $+\frac{\pi}{2}$ or $-\frac{\pi}{2}$.
A qualitative physical explanation is given to this curious fact. We discuss in the final \S \ref{dis.sec}
why our model planet rarely fails to jump through the 2:1 resonance despite the tightness of the
required conditions to do so. We derive the probabilities of capture in spin-orbit resonances for
both prograde and retrograde evolution of spin rate, and compare them with the ranges of
equilibrium eccentricities.

\section{Harmonic expansions of torques}
The instantaneous torque acting on a rotating planet is the sum of the triaxial torque, caused by
the quadrupole inertial momentum, and the tidal torque, caused by the dynamic deformation of
its body. In neglect of the obliquity \citep{danb},
\eb
\ddot \theta=\frac{T_{\rm TRI}+T_{\rm TIDE}}{\xi M_2R^2}
\label{eq.eq}
\ee
with $\theta$ being the sidereal angle of the planet reckoned from the axis of its largest elongation.
All other notations used in this formula and throughout the paper are explained in Table \ref{nota.tab}.
We consider the specific, but representative, case when the obliquity of the planet's equator
is small ($i\simeq 0$) and the planet is not too close to the star ($\frac{R}{a}\ll1$). Neglecting
the precession and nutation of the planet, the triaxial torque is \citep{danb,gold}
\eb
T_{\rm TRI}=-\frac{3}{2}(B-A)n^2\frac{a^3}{r^3}\sin2(\theta-\nu).
\label{tri.eq}
\ee
Using the comprehensive development of Kaula and Darwin's harmonic decomposition of the tidal torque
by \citep{efr2,efrw,efr1,efrb}, one can write a simplified equation for the tidal torque, which we call
in the following Efroimsky's torque
\begin{eqnarray}
T_{\rm TIDE}&=&\frac{3}{2}{\cal G} M_1^2\frac{R^5}{a^6}\sum_{q=-1}^4 G_{20q}(e)
\sum_{j=-1}^4 G_{20j}(e) \nonumber\\  & &
\left[K_c(2,\chi_{220q})\,{\rm Sign}(\omega_{220q})\cos\left((q-j)M\right) +
K_s(2,\chi_{220q})\sin\left((q-j)M\right)\right],
\label{tide.eq}
\end{eqnarray}
where the positively defined forcing frequency
\eb
\chi_{220q}=|\omega_{220q}|=|(2+q)n-2\dot\theta|,
\ee
the so-called G-functions of Kaula $G_{20j}$ are related to power series in eccentricity via Hansen's 
coefficients \citep[e.g.,][]{dobr}:
\eb
G_{20j}(e)=X^{-3\phantom{x}2}_{2+j}(e),
\ee
and the all-important``quality functions" are
\begin{eqnarray}
K_c(l,\chi)&=&-\frac{3}{2(l-1)}\frac{\Lambda_l\chi \Im(\chi)}{(\Re(\chi)+\Lambda_l\chi)^2+\Im^2(\chi)}\\
K_s(l,\chi)&=&\frac{3}{2(l-1)}\frac{\Re^2(\chi)+\Im^2(\chi)+\Lambda_l\chi \Re(\chi)}{(\Re(\chi)+\Lambda_l\chi)^2+\Im^2(\chi)}.
\label{qual.eq}
\end{eqnarray}
Finally, the remaining terms in this equation are
\eb
\Lambda_l=\frac{4\pi(2l^2+4l+3)R^4\mu}{3l{\cal G}M_2^2},
\ee
and
\begin{eqnarray}
\Re(\chi)&=&\chi+\chi^{1-\alpha}\tau_A^{-\alpha}\cos\left(\frac{\alpha\pi}{2}\right)\Gamma(1+\alpha)
\nonumber\\
\Im(\chi)&=&-\tau_M^{-1}-\chi^{1-\alpha}\tau_A^{-\alpha}\sin\left(\frac{\alpha\pi}{2}\right)\Gamma(1+\alpha)
\end{eqnarray}
The $\Re(\chi)$ and $\Im(\chi)$ functions are the real and imagery parts of the complex compliance,
respectively.
Eq. \ref{qual.eq} includes oscillating terms of the tidal torque, $q\ne j$. For a planet similar to
Mercury, these oscillating terms prove very small and can be safely neglected.\footnote{The oscillating components of the tidal torque can not be neglected for a nearly spherical planet, i.e., when
$(B-A)/C$ is very small.} The results of integration over 20000 years were hardly different with or
without these terms. Omitting the oscillating terms of the tidal torque, we arrive at 
the simplified expression
for the secular part of the torque
\eb
T_{\rm TIDE}=\frac{3}{2}{\cal G} M_1^2\frac{R^5}{a^6}\sum_{q=-1}^4 G^2_{20q}(e)
K_c(2,\chi_{220q})\,{\rm Sign}(\omega_{220q}).
\label{sec.eq}
\ee

\begin{deluxetable}{lr}
\tablecaption{Explanation of notations \label{nota.tab}}
\tablewidth{0pt}
\tablehead{
\multicolumn{1}{c}{Notation}  &
\multicolumn{1}{c}{Description}\\
}
\startdata
$\xi$ & \dotfill moment of inertia coefficient \\
$R$ & \dotfill radius of planet \\
$T$ & \dotfill torque \\
$M_2$ & \dotfill mass of planet \\
$M_1$ & \dotfill mass of star \\
$a$ & \dotfill semimajor axis of planet \\
$r$ & \dotfill instantaneous distance of planet from star \\
$\nu$ & \dotfill true anomaly of planet \\
$e$ & \dotfill orbital eccentricity \\
$M$ & \dotfill mean anomaly of planet \\
$B$ & \dotfill second moment of inertia \\
$A$ & \dotfill third moment of inertia \\
$C$ & \dotfill moment of inertia around spin axis\\
$n$ & \dotfill mean motion, i.e. $2\pi/P_{\rm orb}$ \\
${\cal G}$ & \dotfill gravitational constant, $=66468$ m$^3$ kg$^{-1}$ yr$^{-2}$ \\
$\tau_M$ & \dotfill Maxwell time, i.e., ratio of viscosity to unrelaxed rigidity\tablenotemark{\dag} \\
$\mu$ & \dotfill unrelaxed rigidity modulus \\
$\alpha$ & \dotfill tidal lag responsivity \\
\enddata
\tablenotetext{\dag}{For the differences between relaxed and unrelaxed rigidity moduli, cf. \citet{efr1,efrb}.}
\end{deluxetable}

Note that expansions (\ref{tide.eq} and \ref{sec.eq}) are
limited to $l=2$. The higher order expansion terms ($l=3$)
are neglected, because the coefficients are at least 8 orders of magnitude smaller than the $l=2$ terms for
Mercury or similar planets. The expansion in harmonics of mean anomaly is limited to the
range $q,j=-1,0,\ldots,4$ for the ease of computation, the other terms being much smaller than these six.
For example, $G_{205}(0.20563)=0.019$, to be compared with $G_{200}(0.20563)=0.896$, and the rest of 
the Kaula coefficients for the omitted terms are even smaller. The crucial differences
between this approach and the previous studies, e.g. by \citet{cel}, is that more realistic
relations
are employed here for the ``quality functions" of tidal frequency (Eq. \ref{qual.eq}) with 
rheological parameters supported by observations and theory, and that full account is taken of the fast, oscillating terms of the
triaxial torque, which are integrated at a small step.

\begin{deluxetable}{lr}
\tablecaption{Parameters of the Mercury integration \label{data.tab}}
\tablewidth{0pt}
\tablehead{
\multicolumn{1}{c}{Parameter}  &
\multicolumn{1}{c}{Value}\\
}
\startdata
$\xi$ & \dotfill $\frac{2}{5}$ \\[1ex]
$R$ & \dotfill $2.44\cdot 10^6$ m \\
$M_2$ & \dotfill $3.3\cdot 10^{23}$ kg \\
$M_1$ & \dotfill $1.99\cdot 10^{30}$ kg \\
$a$ & \dotfill $5.79\cdot 10^{10}$ m \\
$e$ & \dotfill 0.20563 \\
$(B-A)/C$ & \dotfill $1.2\cdot 10^{-4}$ \\
$n$ & \dotfill $26.088$ yr$^{-1}$ \\
$\tau_M$ & \dotfill 500 yr \\
$\mu$ & \dotfill $0.8\cdot10^{11}$ kg m$^{-1}$ s$^{-2}$\\
$\alpha$ & \dotfill $0.2$ \\
\enddata
\end{deluxetable}

\section{Integration of Mercury}
\label{int.sec}
The evolution of Mercury's spin can be considered as a template model of a rocky, relatively
small planet whose eccentricity can be driven to sufficiently large values due to the interactions
with larger planets in the system. It is also one of the few planets whose physical
parameters are fairly well known \citep{and}. The parameters used in our computations
are listed in Table \ref{data.tab}. The Maxwell time $\tau_M$ was set to the Earth's value,
assuming a similar composition and temperature. The sensitivity of this crucial parameter
to temperature has to be taken into account for rocky exoplanets whose interiors may be
significantly hotter. Higher temperatures result in smaller viscosity and, thus, much shorter
Maxwell times. The $\mu$ parameter was also set to the value for the Earth. The characteristic time
$\tau_A$ is in fact frequency-dependent, and we modeled it as
\eb
\tau_A(\chi)= 50000\,\exp(-\chi/0.2)+500
\ee
in years. This expression is chosen to represent the expected behavior of $\tau_A$,
which sharply rises toward small frequencies. However, the results proved quite insensitive 
to the functional form of $\tau_A$.
In fact, practically the same conclusions can be obtained by simply setting $\tau_A=\tau_M$.
Again, this may not be the case for hotter, less viscous planets.

\begin{figure}[htbp]
\epsscale{1.15}
  \plottwo{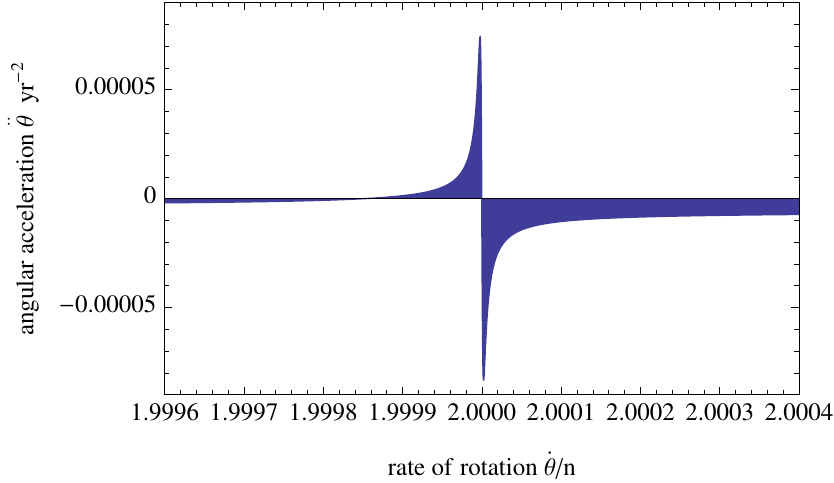}{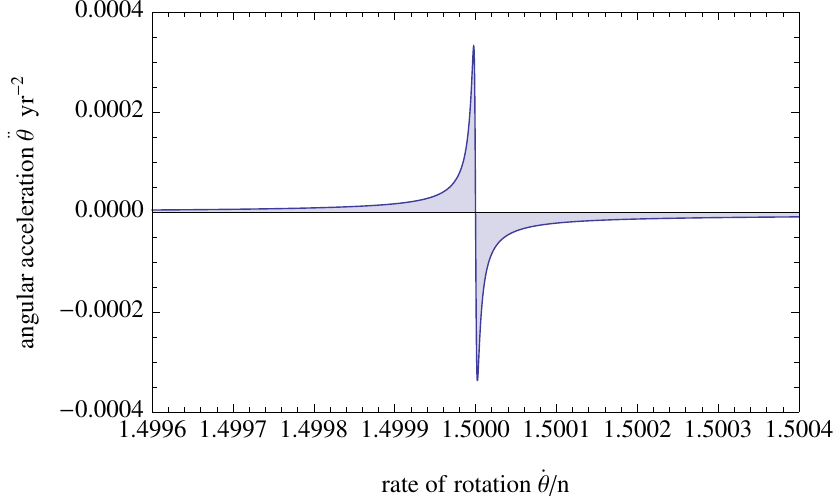}
\caption{Rotation acceleration caused by the secular tidal torque (i.e., $j=q$ in Eq. \ref{tide.eq},
or Eq.~\ref{sec.eq})
on a Mercury-like planet
in the vicinity of the 2:1 resonance (left) and 3:2 resonance (right). Dramatic variations of the torque
are confined to very narrow ranges of frequencies, while the effective
torque integrated over a wider range of frequencies is negative. Despite their appearance on the chosen scale, 
the functions are smooth and differentiable. \label{tide.fig}}
\end{figure}

\begin{figure}[htbp]
  \centering
  \includegraphics{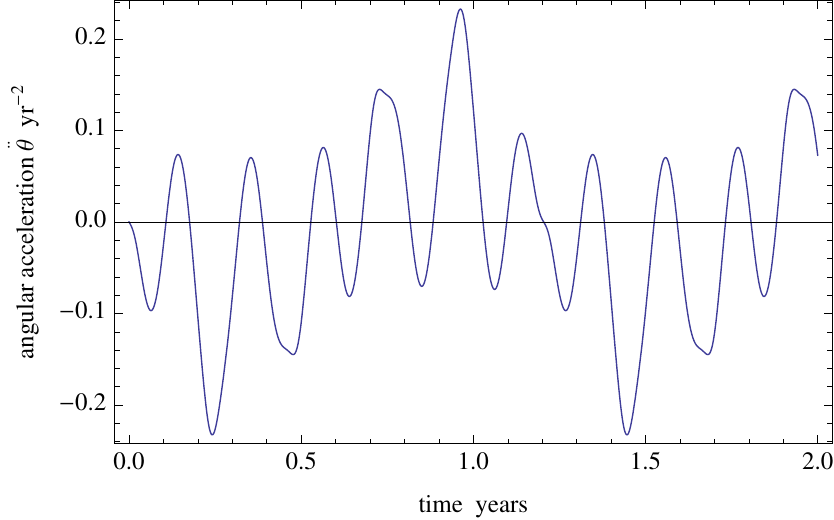}
\caption{Rotation acceleration caused by the triaxial torque on a Mercury-like planet
as a function of time at $\dot\theta=1.6\,n$.\label{tri_acc.fig}}
\end{figure}

The tidal torque at a given eccentricity is a slowly varying function of spin rate everywhere
except the vicinity of spin orbit resonances $\dot\theta=(1+q/2)n$. Fig. \ref{tide.fig}
shows in detail the dependence of the overall angular acceleration $\ddot\theta$ of the planet
caused by the secular tidal torque only, in the vicinity of the 2:1 ($q=2$) and
3:2 ($q=1$) resonances. Note that
the dramatic changes of tidal torque take place within a very narrow interval of tidal
frequencies. The widely used assumption that $\ddot\theta\propto -\dot\theta$ is justified
only in a vanishingly small range of spin rates, i.e., $\dot\theta\in
[1.9999977,2.0000023]\,n$ in the case of 2:1 resonance. The amplitude of the oscillatory
triaxial torques outside the 1:1 resonance is a few orders of magnitude greater than the peak 
values of the tidal
torque (Fig. \ref{tri_acc.fig}). The scale of free librations, caused by the triaxial torque, 
is also a few orders
of magnitude greater than the characteristic width of the kink in Fig. \ref{tide.fig}.
The net effect of the secular
tidal force integrated over one libration period is to spin the planet down. 
This follows from the fact that the mean value of
acceleration at $\dot\theta=2n$ integrated over a much wider interval than the width of the kink,
is negative for these resonances. The peak torque below the resonance rate tapers off quickly 
and becomes negative
at $\dot\theta=1.999895\,n$. The torque function near the 3:2 resonance (Fig. \ref{tide.fig}, right)
is similar in shape,
but the amplitude is larger and the positive shoulder below the resonance is much broader.
The 1:1 resonance is fundamentally different from the higher order counterparts in that the
tidal torque is positive at all rotation rates between 0 and $n$. 

\subsection{Probabilities of capture in 2:1 and 3:2 resonances}
\label{prob.sec}
The ordinary differential equation \ref{eq.eq} was integrated with a grid of initial conditions
$\theta(0)$, $\dot\theta(0)$ for $30\,000$ to $55\,000$ years with a maximum step of $2\cdot10^{-3}$
yr. The initial mean anomaly was always set $M(0)=0$. For the 2:1 resonance, a grid of
40 integrations with $\dot\theta(0)=2.013\,n$ and $\theta(0)=\pi\,i/40$, $i=0,1,\ldots,39$ was performed.
These integrations resulted in 9 captures and 31 passages of the resonance. Therefore,
the estimated probability of 2:1 capture is 0.23 at the current value of eccentricity. The amplitude 
of free libration gradually increases as the planet
spins down toward the point of resonance. Once a lower swing in $\dot\theta$ reaches the resonance, 
a fast transition into
a new spin state occurs on a time scale of a few years. The average rate takes a sudden leap
to significantly smaller value, and the libration starts to decrease. This set of simulations
shows that the chance of Mercury to be captured into the 2:1 resonance at the current value of
eccentricity is modest.

The outcome of integrations in the vicinity of the 3:2 resonance is quite different.
A similar set of integrations with $\dot\theta(0)=1.518\,n$ and $\theta(0)=\pi\,i/40$, $i=0,1,\ldots,39$
resulted in 40 captures. The estimated probability of 3:2 capture is 1.
As soon as $\dot\theta$ reaches the point of resonance, the amplitude
of libration abruptly doubles up, but the mean rotation stays around $1.5\,n$. Once the planet is captured,
the amplitude of libration starts to slowly decline.  The period of libration starts to decrease 
immediately after the capture. A longer integration for $100\,000$ yr shows that the
period of libration asymptotically approaches 16 yr, which is the theoretical value \citep{mur}
for the physical parameters listed in Table \ref{data.tab}. This confirms the validity of
our computations of the triaxial torque. Further numerical experiments revealed
that the planet traverses the 3:2 resonance in a similar manner to the 2:1 resonance at
significantly smaller values of eccentricity. We conclude that our numerical simulations are
consistent with the fact that Mercury is entrapped in the 3:2 resonance with the current
value of eccentricity. This outcome is indeed most likely unless the eccentricity acquired
much different values in the past. An even higher eccentricity would have resulted in the entrapment
in the 2:1 resonance long in the past when the planet was approaching this rotation rate. Alternatively,
a considerably smaller eccentricity at the moment of approaching the 3:2 resonance would have made
the planet traverse it quickly and continue to spin down.

\subsection{Equilibrium torques and evolution of eccentricity}
\label{equi.sec}
\citet{corla04} pointed out that Mercury's eccentricity has varied chaotically during its long
dynamical evolution in the Solar system. The current state of Mercury, entrapped in the 3:2
spin-orbit resonance, is the result of a long history of tidal and orbital interactions, probably
marked by multiple passages of spin-orbit resonances. In particular, if the initial spin
rate of Mercury was much greater than it is today, the planet has successfully traversed
a number of higher resonances. Understanding the dependence of tidal torque on eccentricity in
the framework of Efroimsky's model is as important as the truthful estimation of capture
probabilities in each resonance.

\begin{figure}[htbp]
  \centering
  \includegraphics[angle=0,width=0.9\textwidth]{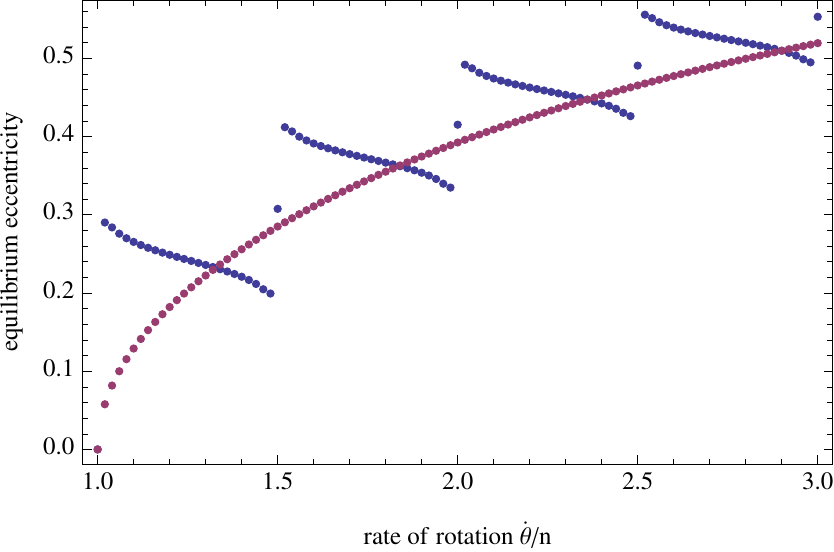}
\caption{Equilibrium eccentricity of a Mercury-like planet separating the areas of negative (spin-down)
and positive (spin-up) secular tidal torque for a grid of relative rates of rotation.
The smooth curve represents the linear torque widely used in the literature. The jagged
curve represents Efroimsky's torque.\label{equi.fig}}
\end{figure}

\citet{corla04,corla09} employed a linear torque model, also known as MacDonald's torques \citep{gold}
or Constant Phase Lag model \citep{fer}, in which the secular torque is linearly dependent on
the time derivative of tidal frequency. As we noted in \S \ref{int.sec}, Efroimsky's torque is
linear only in vanishingly narrow intervals around resonances (Figs. \ref{tide.fig}a and b).
The widely used assumption that the secular tidal torque is linear everywhere across the
range of relative spin rates $\dot\theta/n$ leads to well-known problems and inconsistencies.
For example, a slowing down Moon in the absence of quadrupole momentum 
is not allowed to descend into the 1:1 resonance (synchronous rotation) for
any nonzero eccentricity \citep{mur,wil}, because the linear torque,
monotonously increasing with growing eccentricity, changes sign from negative to positive
at a certain equilibrium eccentricity for any fixed $\dot\theta/n > 1$. Therefore, a perfectly
spherical Moon should have stalled
in its de-spinning at a higher than synchronous rate, namely, $\dot\theta_{\rm equ}/n = 1+6e^2=1.018$.
This is not the case for Efroimsky's tidal torque, which sharply decreases at supersynchronous
rotation rates, counteracting the effects of eccentric motion.
The dependence of equilibrium eccentricity, whence the linear torque disappears, for a range of
rotation rates is depicted in Fig. \ref{equi.fig} with the monotonously rising dotted line. It implies
that synchronous rotation, so commonly observed among the satellites of the Solar System, is
not attainable for any nonzero eccentricity. If Mercury traversed the 3:2 resonance and continued to
spin down toward synchronization, it would be stuck at an equilibrium rate of $1.24\,n$ with the current
value of eccentricity.

The character of equilibrium torques is profoundly different with Efroimsky's model, depicted with
the jagged dotted curve in Fig. \ref{equi.fig}. The curve was obtained by finding the
roots of Eq. \ref{sec.eq} in $e$ for a grid of $\dot\theta/n$. The first jump from $e_{\rm equ}=0$
at $\dot\theta/n = 1$ to $e_{\rm equ}\approx 0.29$ just slightly above the synchronous rate
implies that Mercury is allowed to be captured in the 1:1 resonance if the eccentricity is not
too large ($e<0.29$), evolving either from slower or faster rotation rates. However, for faster
rotation rates between $1\,n$ and $1.5\,n$, Mercury can stall in its de-spinning unless $e<0.2$.
Likewise, entrapment in the 3:2 resonance is inevitable for $0.2<e<0.41$ if Mercury reaches
this rate of rotation from either direction. This is consistent with the results of numerical simulations
in \S \ref{prob.sec} that the probability of 3:2 capture with the current value of eccentricity is 1.
What happens if Mercury reaches $\dot\theta/n = 1.5$ with an eccentricity below 0.2? Capture
is still possible, but the probability declines with decreasing eccentricity, cf. \S \ref{dis.sec}.
Entrapment of the planet in the 2:1 resonance is inevitable for $0.34<e<0.49$, etc. The teeth
of the equilibrium torque curve act as very efficient resonance traps for a planet like Mercury,
whose eccentricity varies in a fairly wide range over billions of years of dynamical evolution.

\section{Conditions of traversing a resonance}
\label{con.sec}
\begin{figure}[htbp]
  \centering
  \includegraphics[angle=0,width=0.7\textwidth]{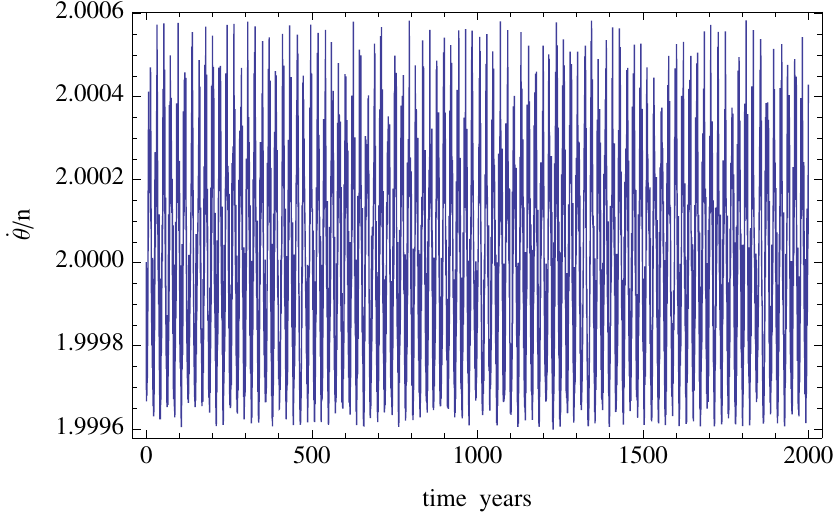}
\caption{Results of integration of the Mercury-like test planet with initial conditions
of resonance, $\dot\theta(0)=2\,n$, $\theta(0)=0$. The spin rate remains in resonance to
the orbital motion indefinitely long. \label{stay.fig}}
\end{figure}

The purpose of our next set of simulations is to find out under which circumstances the
planet traverses a spin-orbit resonance. We have established that the test planet passes
the point $\dot\theta=2\,n$ quite quickly, within several years. At some moment during
this passage, the planet is at the periastron with a rotation rate close to $2\,n$. What
is the rotation angle $\theta$ at this time, and does this value matter for the way
this dynamical transformation unfolds? One hundred integrations were conducted
with the planet initially already at resonance spin rate ($\dot\theta(0)=2\,n$), but with
initial phase angles ranging from $-\pi/2$ to $+\pi/2$ in equal steps. 
For most of the
interval of possible initial angles, the planet is clearly captured
in the resonance, i.e., the spin rate continues to oscillate around the resonant value. A typical example
of such entrapment with the initial conditions $\dot\theta(0)=2\,n$, $\theta(0)=0$, is shown
in Fig. \ref{stay.fig}. The rate of rotation oscillates around the resonant point roughly
between $1.9996\,n$ and $2.0006\,n$, seemingly forever. The oscillations (not resolved in the
Figure) are distinctly non-sinusoidal immediately after the capture.
The amplitude of these oscillations slowly declines with time. The tidal bulge still runs across
the circumference of the planet with a period equal to the orbital period; therefore the tidal
dissipation of energy goes on. A slow shrinkage of the orbit is the main
source of tidal energy for a planet captured in spin-orbit resonance. 

\begin{figure}[htbp]
  \centering
  \includegraphics[angle=0,width=0.7\textwidth]{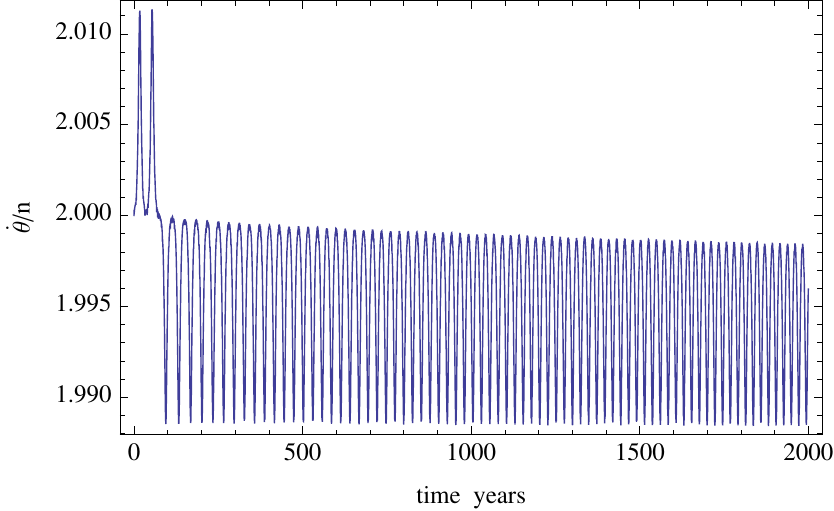}
\caption{Similar to fig. \ref{stay.fig}, but with a different initial phase
angle, $\theta(0)=\frac{\pi}{2}$. The planet traverses the resonance after a couple
of libration oscillations, in less than 100 yr. \label{escape.fig}}
\end{figure}

The result is drastically different if we set the initial conditions to $\dot\theta(0)=2\,n$, $\theta(0)=
\pi/2$ (Fig. \ref{escape.fig}). After a couple of upward swings, the spin rate jumps through the resonance point in $\sim 80$ years. The spin rate resets abruptly at considerably lower mean values.
The peak rates never quite reach the resonance value, gradually diminishing with time. A sidereal
angle $\theta=\pm \pi/2$ at $M=0$ implies that the planet is positioned sidewise with respect to
the star, that is, its longer dimension is perpendicular to the Sun-planet line. This is
the preferred configuration, which is necessary for the planet to traverse the resonance. The range of suitable phase angles at $\dot\theta(0)=2\,n$ and $M(0)=0$, which make it possible for the
planet to traverse the resonance, is quite small,
approximately between $\frac{\pi}{2}-0.030$ and $\frac{\pi}{2}+0.021$.
Remarkably, at any other $\theta(0)$ in $[0,\pi]$, the planet remains entrapped in this resonance.
Thus, the passage through resonance can not occur unless the planet is turned almost sidewise with 
respect to the star at one of the periastrons. A stronger tidal dissipation during the last
circulation before the resonance would not allow the planet to reach this range, and capture becomes inevitable.

Such strong requirements to the orientation of the planet at the point of resonance may
seem puzzling at first glance. However, simple qualitative considerations of the interaction
of the tidal and triaxial torques explain the matter. Let us consider a triaxial planet in a
resonance $2\dot\theta=(2+q)n$, so that for each 2 orbital revolutions, it makes exactly $2+q$ rotations about
its axis. Assume $\theta(0)=0$
at an initial periastron passage, so that the longest dimension is aligned with the 
instantaneous direction to the star. After two complete orbital revolutions,
the planet arrives at the periastron with a slight lag in phase angle, because the tidal torque
decelerated the planet's rotation. The planet therefore is tilted opposite to the spin direction.
The triaxial torque will act to re-align the planet again,
so that the average action from the triaxial torque is to spin up the planet again. This counter-action
of the triaxial torque is symmetric, in that if the planet's spin accelerated during
the two complete orbits the torque
will rectify its rotation. Thus, any deviation of spin rate from the resonance value will be
automatically corrected by the triaxial torque. The planet is trapped in a stable equilibrium.
This is not the case when the initial angle at periastron is $\pm\pi/2$.
Any net lag in phase angle will cause a nonzero triaxial torque in the same direction, causing the
planet to continue to spin down. In this configuration of unstable equilibrium, the torque assists the
tides to swiftly turn the planet around and lunge through the resonance.

\begin{figure}[htbp]
  \centering
  \includegraphics[angle=0,width=0.7\textwidth]{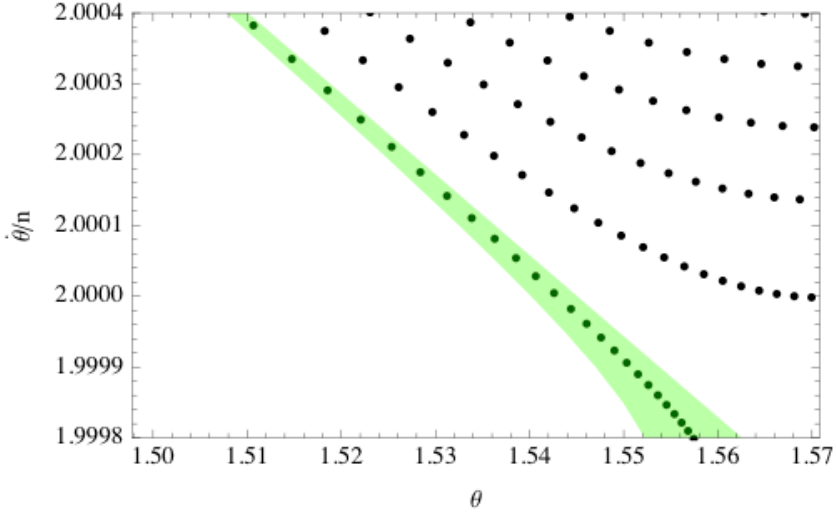}
\caption{A small section of the $\{\theta,\dot\theta\}$ phase space around the
2:1 spin-orbit resonance. The green-shaded area is the subspace of initial parameters,
from which a direct passage through resonance takes place. The dots show consecutive
periastron positions of the test planet. Parts of four libration swings gradually approaching
the point of resonance are visible before the final
trajectory hits the passage area and follows it through the resonance into the
domain of lower spin rates, $\dot\theta<2\,n$. \label{needle.fig}}
\end{figure}

The same mechanics work for a circular orbit ($e=0$). One may ask then, how a planet
initially at $\theta(0)=0$ can be entrapped at all, if at some
point afterwards it inevitably turns sidewise with respect to the star?
The answer is that while the planet turns through $\pi/2$ with respect to the star, the spin rate
will change because of the triaxial torque, and will no longer be equal
to $(1+q/2)n$. In other words, the condition of resonance
passage is a certain area of the 2D phase space.\footnote{In fact, the relevant phase space is
three-dimensional if we include the initial mean anomaly $M(0)$, but for simplicity of analysis,
we restricted our study to the plane $\{\theta,\dot\theta\}$ by considering only the
instances of periastron.} Returning to Fig. \ref{escape.fig}, we note that the point of
the phase space $\{\theta(0)=\pi/2,\dot\theta(0)=2\,n\}$ does not actually belong to this
area, because the passage through resonance is not immediate. Indeed, the planet has to reallign
its phase space parameters in two upward swings before it traverses the resonance at about 80 yr
after the start of integration. In order to map the phase space of {\it immediate} resonance passages,
we performed several hundred short-term integrations (for 1000 yr) varying the initial $\dot\theta(0)$
with a step of $5\cdot10^{-5}n$ and $\theta(0)$ with a step of $1\cdot10^{-5}$ rad. The mapped section of
this space for our Mercury-like planet and the 2:1 resonance is shown in Fig. \ref{needle.fig}
as green-shaded area. If the planet at periastron turns up within the green area,
it immediately traverses the resonance. This "green corridor" is extremely narrow in $\theta$
at $\dot\theta/n>2$ and occupies a tiny fraction of the phase space. The black dots display the
periastron positions of the planet. The dots are lined up in five consecutive libration trajectories,
each following one probing lower spin rates around $\theta=\pi/2$. The final lap hits the ``needle's eye"
and falls through the resonance, never to return to it. Note that the last but one trajectory actually
reached $\dot\theta=2\,n$ at $\theta$ very close to $\pi/2$, but the tipping point is at slightly
smaller $\theta$. 

\section{Discussion}
\label{dis.sec}
The area of the phase space above the resonant spin rate, through
which a Mercury-like planet can traverse a resonance, is very narrow. For example,
the appropriate phase angle at periastron for $\dot\theta=2\,n$ should be between
$\pi/2-0.03065$ and $\pi/2-0.02575$. It may appear improbable that the planet, driven by the
relatively large triaxial torques, and wandering through good part of the phase space in its
pre-resonance evolution, can hit this very small passage opening. And yet, as we established through
many numerical integrations, the test planet in most cases passes the resonances higher
than 3:2, if the integration starts well above the resonant spin rate. The camel rarely fails to
go through this needle's eye. 

\begin{figure}[htbp]
  \centering
  \plottwo{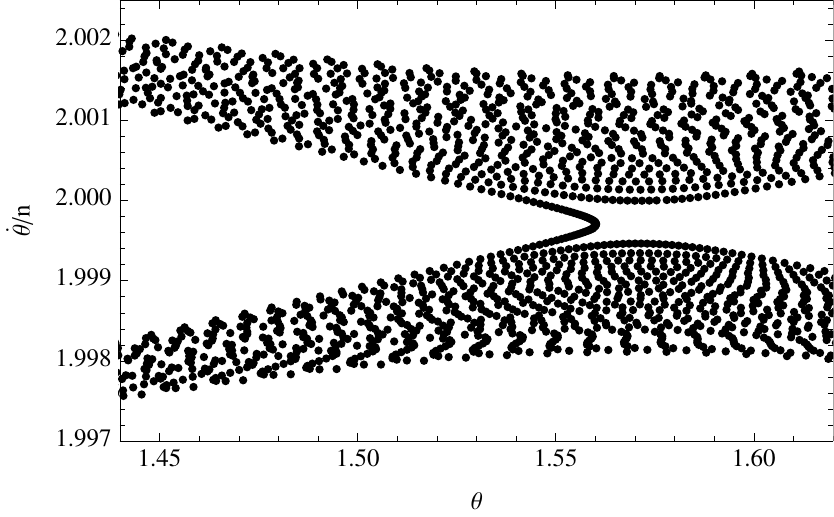}{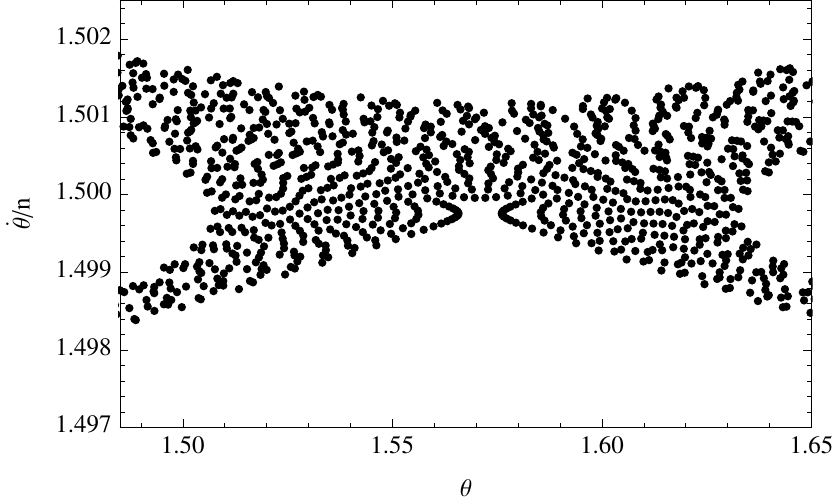}
\caption{Phase space position at times of periastron of a planet passing through the 2:1
resonance (left) and captured in the 3:2 resonance (right). \label{phase.fig}}
\end{figure}

In the domain of above-resonance spin rates, the passage area extends almost along a straight line
toward higher $\dot\theta/n$ and smaller $\theta$. Fig. \ref{needle.fig} shows only a segment
of this area. The needle's eye area looks more like a needle, stretching out to at least
$\theta=\pi/2-0.9031$, $\dot\theta/n=2.009$, tapering off to a point. On the other hand,
the lower extents of libration swings tend to probe the area of unstable equilibrium, i.e., 
the minimum point of each trajectory is close to $\theta=\pi/2$. This important fact
follows from the integral of energy for free librations in the vicinity of a resonance $q$
averaged over one orbital period
\citep{mur}, which can be written as
\eb
\frac{1}{4}C\,\dot\gamma^2-\frac{3}{2}n^2(B-A) G_{20q}(e)\cos(\gamma)=E_0
\ee
where $\gamma=2\theta-(2+q)M$ and $E_0$ is a constant energy. When the rate of rotation is still faster than
the resonant value, $2\dot\theta> (2+q)n$, but it is at the minimum of a libration swing, $\dot\theta=$min,
the cosine in this equation should be equal to $-1$, hence, $\gamma=\pi$. At the times of periastron, the corresponding value of $\theta$ is
$\pi/2$. Figs. \ref{phase.fig}a and b show larger parts of the phase space in the vicinity of
a 2:1 resonance passage and a 3:2 resonance capture, respectively. In the case of passage, the trajectory
lunges through the opening depicted in Fig. \ref{needle.fig} from the upper zone of circulation to the lower zone of circulation, reversing its direction. In the case of capture, the
trajectory breaches the separatrix between the upper circulation zone and the central zones of pure libration
where it follows gradually tightening loops.

The analogy with a rotating pendulum by \citet{gol68} can help to qualitatively understand
the alignment required for the planet to traverse the resonance. The pendulum is swinging over 
the top of its support, while a secular torque is
slowing down its rotation. Inevitably, the pendulum looses enough rotational energy to be unable to
pass the highest point and for a while stops close to the highest point, after which it starts to rotate in the opposite direction. But the secular torque acts in the same direction, this time assisting the pendulum
in passing the top in the counter direction. Once the average $\dot\gamma$ changed sign, the pendulum
continues to rotate in the counter direction with acceleration. This also explains why
the trajectories in Fig. \ref{needle.fig} are so flat at the minima of the libration oscillations.
Descending step by step to lower spin rates at each libration swing, the trajectories
can hardly avoid catching
on the nearly linear area of immediate passage through the resonance.

\begin{figure}[htbp]
  \centering
  \plottwo{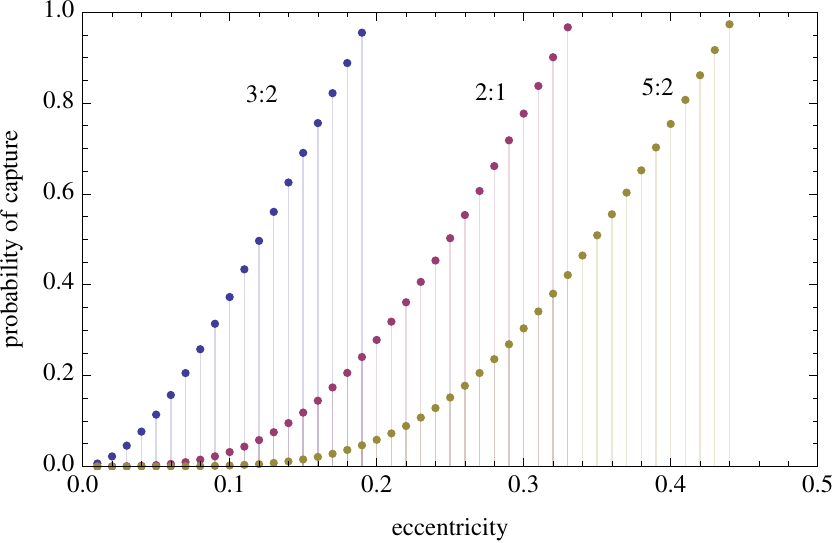}{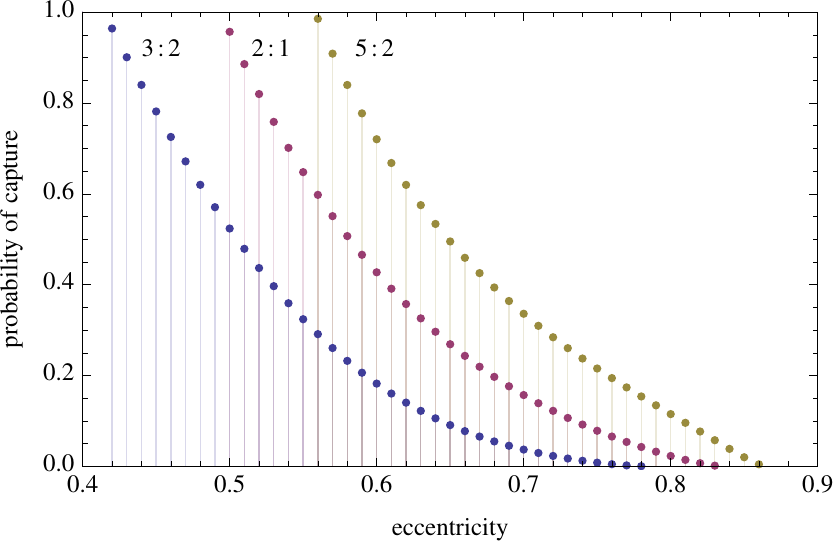}
\caption{Probability of capture of a Mercury-like terrestrial planet in
3:2, 2:1 and 5:2 spin-orbit resonances. Left: prograde evolution when the tidal torque
is generally negative and the planet is spun down. Right: retrograde evolution when the tidal torque
is generally positive and the planet is spun up.\label{prob.fig}}
\end{figure}

This remarkable fact can be further elucidated by analysis of capture probabilities following the
lines in \citep{gold}. First, we note that the secular tidal torque in the vicinity of a
resonance $q'$ (Eq. \ref{sec.eq}) can be split into two parts, one including the $q=q'$
term and the other the rest of the sum. The $q=q'$ term is an odd function of $\omega_{220q'}$
around zero tidal frequency. The other term, which we call bias, is to a very good approximation
constant with $\omega_{220q'}$, because it is the sum of all other secular torques at resonances
outside $q'$. These resonances are spaced by $n$ in $\dot\gamma=-\omega_{220q'}$, whereas the amplitude
of librations close to a resonance is much smaller (Figs. \ref{stay.fig} and \ref{escape.fig}). Figs. 
\ref{tide.fig}a and b show that the bias is negative for the given eccentricity, implying a 
(nearly) frequency-independent
dissipation of rotation energy. It is sufficient to consider two librations around the point
of resonance $\dot\gamma=0$, i.e., the last libration with positive $\dot\gamma$ and the first
libration with negative $\dot\gamma$. \citet{gold} noted that if the energy offset from zero
at the beginning of the last libration above the resonance is uniformly distributed between
$0$ and $\Delta E=\int\langle T\rangle\dot\gamma dt$, where $\langle T\rangle$ is the
secular torque, the probability of capture is
\eb
P_{\rm capt}=\frac{\delta E}{\Delta E},
\ee
with $\delta E$ being the total change of kinetic energy at the end of the libration below
the resonance. Thus, $\langle T\rangle\dot\gamma$ should be integrated over one cycle
of libration to obtain $\Delta E$, and over two librations symmetric around the resonance $\omega_{220q'}
=0$ to obtain $\delta E$. As a result, the odd part of the tidal torque at $q=q'$ doubles in
the integration for $\delta E$, whereas the bias vanishes. For the secular torque in Eq.
\ref{sec.eq}, using the singular separatrix solution of zero energy\footnote{Note that our $\gamma$
is twice the $\gamma$ in \citep{gold}}
\eb
\dot\gamma=2\,n\,\left[ \frac{3(B-A)}{C}G_{20q'}(e)\right]^\frac{1}{2}\cos\frac{\gamma}{2},
\label{separ.eq}
\ee
we obtain
\eb
P_{\rm capt}=\frac{2}{1+2\pi V/\int_{-\pi}^{\pi}W(\dot\gamma)d\gamma}
\label{prob.eq}
\ee
where
\begin{eqnarray}
V&=&\sum_q G^2_{20q}(e)K_c(2,|q-q'|n)\,{\rm Sign}(q-q')\nonumber\\
W(\dot\gamma)&=& -G^2_{20q'}(e)K_c(2,\dot\gamma)
\end{eqnarray}
We included the term $q=q'$ for bias $V$ {\it pro forma}, because $K_c(2,0)=0$. The integral
in Eq. \ref{prob.eq} can be computed numerically using Eq. \ref{separ.eq}. The results of this semi-analytical
estimation of capture probability as a function of eccentricity are presented in Fig. \ref{prob.fig}, left.
It is gratifying to see that they are in agreement with the results of brute-force
simulations discussed in \S \ref{prob.sec}. At $e=0.205$, the probability of capture
in the 3:2 resonance is 1, and the probability of capture in the 2:1 resonance is approximately $0.3$.

This fast way of computing capture probabilities can be applied to exoplanets of terrestrial
composition, keeping in mind that the probabilities depend on the degree of triaxiality through the
parameter $(B-A)/C$ in the equation of separatrix (Eq. \ref{separ.eq}). Even the smallest
exoplanets discovered to date tend to be larger than the Earth because of an observational selection
effect. Larger rocky planets are likely to be more axially symmetric. For smaller $(B-A)/C$, keeping all
other parameters the same, the curves of capture probabilities in Fig. \ref{prob.fig} become
steeper, so that guaranteed capture is achieved at smaller minimal eccentricities. This is to be
expected, because if the triaxial torque is turned off in Eq. \ref{eq.eq}, capture is inevitable
at any resonance where the tidal torque changes sign. For example, the minimal eccentricities
of inevitable capture of a Mercury-like planets with $(B-A)/C=1.2\cdot 10^{-7}$ drop to $0.08$
for 3:2, $0.20$ for 2:1, and $0.30$ for 5:2 resonances.

\citet{wie} provided observational evidence that Mercury was previously captured in
a synchronous rotation. The authors of this paper explain that this possibility is achievable if
the planet starts its evolution with a retrograde rotation. The secular tidal torque in
this case spins the planet up, first making the rotation prograde, and then driving the
planet toward the 1:1 resonance. Once captured in synchronous resonance, Mercury could
remain there for an extended period of time, until a fortuitous large impact drove it
out of the resonance abruptly increasing its rate of rotation. Under circumstances, the planet
could then cross the higher resonances in the upward order. Since the previous consideration
of capture probabilities via energy balance is completely symmetric with respect to
the direction of $\dot\gamma$, Fig. \ref{prob.fig}, left, is valid for the reverse crossing of
resonances for eccentricities below the equilibrium values. However, more stringent conditions 
of such retrograde evolution come from the consideration
of equilibrium torques in \S\ref{equi.sec}. As follows from Fig. \ref{equi.fig}, a strong impact
would not be sufficient for Mercury to leave the 1:1 resonance. The orbital eccentricity
at the time should be greater than 0.29 to overcome the first barrier of equilibrium torque.
Furthermore, the eccentricity should remain above 0.20 for a continuous spin-up to the point
of 3:2 resonance. There, Mercury is guaranteed to be captured at any eccentricity
between 0.20 and 0.41, because the planet would not be able to spin up any further, whereas the probability
of capture at $e>0.41$ begins to decline with growing eccentricity. Fig. \ref{prob.fig}, right,
shows the probabilities of capture in the 3:2, 2:1 and 5:2 resonances for a retrograde evolution
of spin rate, and for eccentricities exceeding the upper limits of the equilibrium torque.\footnote{
Computation of capture probabilities for retrograde evolution by Eq. \ref{prob.eq} is technically
difficult because the Kaula functions vary rapidly at high eccentricities. It is necessary to include
more terms in the sum in Eq. \ref{sec.eq}, e.g., $q=-2,-1,\ldots, 8$.}
Thus, Efroimsky's tidal model described in this paper does not
rule out the hypothesis by \citet{wie}, but requires, beside the external action, fairly
high values of orbital eccentricity during the ascent to the current 3:2 resonance.

The computations presented in this paper for Efroimsky's model of tidal torque serendipitously resolved 
the long-standing conundrum of Mercury's capture
into the 3:2 resonance. The previous approximations of tidal torque \citep{gol68,mur}
either predicted a low probability of this capture with the current value of eccentricity,
or were not able to reproduce the libration damping. Our initial computations imply
that the probability of entrapment of Mercury in the 3:2 resonance with the current
value of eccentricity is 100\%. No other outcome would have been possible unless the eccentricity acquired
much smaller values in the past, or Mercury's initial rotation was retrograde. An higher 
eccentricity could have likely resulted in the entrapment
in the 2:1 resonance long in the past when the planet was approaching this rotation rate. 

\acknowledgments
I thank the USNO Editorial Board for helpful suggestions and
a critical reading of the original version of the
paper. Dr. M. Efroimsky inspired this paper and generously shared his insight in the
mechanics of tides.

\end{document}